\documentstyle[aps,prl]{revtex}
\twocolumn
\draft
\begin{document}
\input{epsf}
\twocolumn[\hsize\textwidth\columnwidth\hsize\csname@twocolumnfalse\endcsname
\title{
Parametrically forced sine-Gordon equation
and domain walls dynamics in ferromagnets
}
\author{Vadim~Zharnitsky}
\address{Division of Applied Mathematics, Brown University, Providence,
Rhode Island 02912}
\author{Igor~Mitkov}
\address{Applied Theoretical and Computational Physics Division
and Center for Nonlinear Studies\\
Los Alamos National Laboratory, Los Alamos, New Mexico 87545}
\author{Mark~Levi}
\address{
Department of Mathematical Sciences, Rensselaer Polytechnic Institute,
Troy, New York 12180}
\date{\today}
\maketitle

\begin{abstract}
A parametrically forced sine-Gordon equation with a fast periodic
{\em mean-zero} forcing is considered.
It is shown that $\pi$-kinks represent a class of solitary-wave solutions
of the equation. This result is applied to 
quasi-one-dimensional ferromagnets with an easy plane anisotropy, in a
rapidly oscillating magnetic field.
In this case the $\pi$-kink solution we have introduced corresponds
to the uniform ``true'' domain wall motion, since the magnetization
directions on opposite sides of the wall are anti-parallel.
In contrast to previous work, no additional anisotropy is required
to obtain a true domain wall.
Numerical simulations showed good qualitative agreement with the theory.

\end{abstract}
\pacs{
75.60.Ch
03.40.Kf
}
\vskip1pc]
\narrowtext

The sine-Gordon equation (SGE) arises in various physical applications
including Josephson junction transmission lines~\cite{scott,mineev1},
dislocations in
crystals~\cite{frank}, charge density waves~\cite{bishop2}, waves in
quasi-one-dimensional ferromagnetic 
materials~\cite{leung1,leung2,mikeska1,bishop1}.  
The only stable traveling wave solutions of ordinary SGE for a scalar field 
$\phi$ are localized solutions with identical boundary conditions $\phi=0$ and 
$\phi=2\pi$, called $2\pi$-kinks. On the other hand, non-localized kinks 
separating regions with different values of the field (e.g. $\pi$-kinks) are 
also important since they are easier to follow experimentally.  This 
is especially relevant for domain wall dynamics in 
quasi-one-dimensional ferromagnets with strong anisotropy of a hard 
magnetization axis~\cite{leung1,leung2,mikeska1,abdul,kirova}. 
In the presence of an external  magnetic  field (weak compared to the 
anisotropy) applied within the 
easy plane, the spin dynamics can be described by SGE 
(see, {\it e.g.}, 
\cite{leung2}) with $2\pi$-kinks as the only traveling wave solutions.
These localized solutions, preserving the average magnetization, were observed 
experimentally~\cite{steiner,jough}. 
An interesting effect of  evolution of the average magnetization 
takes place if adjacent domains have anti-parallel
magnetizations which correspond to $\pi$-kinks
or ``true'' domain walls. Such $\pi$-kinks were
obtained  by introducing additional anisotropy within the easy 
plane (see, {\it e.g.}, \cite{leung2} and \cite{kosevich}, sect. 8.3).

In this letter we obtain $\pi$-kinks in a modified SGE with a fast mean-zero
parametric excitation. For quasi-one-dimensional ferromagnets 
with an easy plane anisotropy, this implies the existence of true domain walls 
generated by a rapidly oscillating external magnetic field. As opposed 
to~\cite{leung2,kosevich},
no additional anisotropy of the ferromagnet is required for the
existence of $\pi$-kinks.

Various cases of parametrically forced SGE  have arisen in numerous models of
physical systems~\cite{mineev1,kirova,kivshar1,kivshar2}.
However, the mean-zero case has been left out of consideration, while it is
significant both for theory and applications. Indeed,
a zero-average periodic forcing implies that the states
$\phi=0$ and $\phi=\pi$ are symmetric. This leads to
the existence of $\pi$-kinks with such time dependent
macroscopic quantity as the spatial average of the field.

We start with the Hamiltonian~\cite{leung2,bishop3}
\begin{equation}
{\cal H} = \sum_{i=1}^N \left[\,-\; J\,
{\mathbf S}_i \cdot {\mathbf S}_{i+1}
\;+\; D\,(S_i^z)^2 \;-\; g\mu_B H S_i^x\;\right]\;,
\label{igor1}
\end{equation}
where ${\mathbf S}_{i}$ are dimensionless classical spin vectors,
$N$ is the number of spins, and $J$ is a constant of exchange interaction.
The external magnetic field ${\mathbf H}$ is directed along the $X$-axis.
The anisotropy constant $D$ provides the existence of an easy plane,
$XY$, at each site of the chain.

The dynamics of a spin ${\mathbf S}_i$ in the effective magnetic
field is governed by~\cite{kosevich}
\begin{equation}
\hbar \dot{\mathbf{S}}_i \;=\; -{\mathbf S}_i \,\times\,
\frac{\partial {\cal H}}{\partial{\mathbf S}_i}\;.
\label{igor3}
\end{equation}

Representing spins in spherical coordinates,
${\mathbf S}_i = S\,(\cos\theta_i \cos\phi_i,\,\cos\theta_i \sin\phi_i,\,
\sin\theta_i)\,$, and taking the continuum limit,
yields a system of two first-order (in time) partial differential equations
for $\theta$ and $\phi$~\cite{leung2,bishop3}.
When the condition $DS \gg g\mu_{B}H$ holds~\cite{leung2},
the system reduces to SGE for the polar angle in the easy plane, $\,\phi$
\begin{eqnarray}
\phi_{xx} \;-\; \frac{\ddot{\phi}}{C^2} \;=\; \nu \, H\,\sin\phi\;,
\label{igor7}
\end{eqnarray}
where
$C \,=\, 2a_{0}S^2 \sqrt{JD}\,/\,\hbar\,,\;
\nu \,=\, g\mu_B / 2JSa_0^2\,$, and $a_0$ is the 
lattice constant.
We take the external field $H\,=\,H_0\, a(t/\epsilon)$, where
$a$ is a mean-zero periodic function
with a unit amplitude,
and $H_0$ is the amplitude of the field. Rescaling time and coordinate
to dimensionless variables, $\;\tilde{t} \,=\, t\,C\,\sqrt{\nu H_0}\,,
\;\tilde{x} = x\,\sqrt{\nu H_0}\;$,
we obtain (after dropping tildes) a parametrically forced SGE (PSGE)
\begin{equation}
\phi_{tt} \;-\; \phi_{xx} \;+\; a(t/\epsilon)\,\sin \phi \;=\; 0\;.
\label{igor9}
\end{equation}

After averaging Eq. (\ref{igor9}) directly, over the fast time scale 
$\epsilon$, we are left with a linear wave equation with wave propagation
velocity $c=\pm 1$. In what follows, we use more subtle averaging to
obtain $\pi$-kink solutions moving with any prescribed velocity.

The phenomenon of $\pi$--kinks  in PSGE has a finite--dimensional
counterpart: it is the stabilization of the inverted pendulum by periodic
vibration of its suspension point (Kapitza pendulum \cite{landau1}). 
Since this latter phenomenon is responsible
for the existence of $\pi$--kinks, we outline a very simple geometrical
explanation of stability of the inverted pendulum with a vibrating suspension.
Full details  with other applications can be found in \cite{levi}
along with further references to numerous papers on the subject. Consider a
pendulum (a bob on a massless rod of length $l$) in a nearly upside-down 
position, with the  
suspension point undergoing vertical vibration. We assume the latter to have 
high acceleration and small amplitude.   Since the acceleration is large, the
force of the rod on the  bob is large so that  the  bob would be expected
to follow, in the first approximation, the direction of the rod. This suggests
considering an auxiliary system where the velocity of the bob is  actually
constrained to the line of the rod. In this case the bob will oscillate along
an arc  of a tractrix (the "pursuit" curve: all tangent segments from this
curve to a straight line have the  same length) and thus will be a subject to
an average centrifugal force
$ m k\langle v^2 \rangle $, where $k$ is the curvature of the tractrix and $v$ 
is the speed of the bob which is approximately the speed of the suspension 
point
when the pendulum is near the top. If we now release the constraint, thus
releasing the centrifugal force, the bob will  behave as if it were subject
to a centripetal force $m k \langle v^2 \rangle $ which pushes the pendulum
towards the top. If this force exceeds the gravitational force, the pendulum is
stable; this leads to the simple stability criterion $ \langle v ^2 \rangle >
\ell g $, see \cite{levi}.

To average Eq. (\ref{igor9}) we apply a series of canonical 
near-identical transformations via
the normal form technique \cite{arnold,neistadt},
so as to bring the original equation to a better form 
with rapidly oscillating 
coefficients moved to higher order terms. Since the
transformations are near-identical the solutions for the reduced 
Hamiltonian are close to those for the original one. 

The Hamiltonian of PSGE (\ref{igor9}) is given by 
\begin{eqnarray}
H=\int_{-\infty}^{+\infty}\left(\frac{p^{2}}{2}+
\frac{\phi_{x}^{2}}{2}-a\cos{\phi}\right)dx\;,
\end{eqnarray}
where $p\,\equiv\,\phi_t$ (below we omit the limits of integration and $dx$).
Let the first canonical transformation be defined implicitly as follows
\begin{eqnarray} 
p=p_1+\frac{\delta W_{1}(\phi,p_1)}{\delta \phi}\;,\;\;\;\;
\phi_1=\phi+\frac{\delta W_{1}(\phi,p_1)}{\delta p_1}\;.
\label{frsttransf} 
\end{eqnarray}
The new Hamiltonian is given by $H_1=H+W_{1t}$, or
\begin{eqnarray}
H_1=\int
\left[ \frac{1}{2}\left(p_1+\frac{\delta W_{1}}{\delta \phi}\right)^{2}
+\frac{\phi^{2}_x}{2}-a\cos{\phi}\right]  +W_{1t}\;.
\end{eqnarray}
To kill the rapidly oscillating term in $H$ we choose 
$W_1 =\epsilon \int a_{-1}\cos{\phi} $, where 
$a_{-1}$ is an anti-derivative with zero average.  The Hamiltonian 
takes the form
\begin{eqnarray}
H_{1}=\int \frac{p_{1}^{2}}{2}+\frac{\phi_{1x}^{2}}{2}- 
\epsilon a_{-1}p_{2}\sin{\phi_{1}}+
\frac{1}{2}\epsilon^{2}a_{-1}^{2}\sin^{2}{\phi_{1}}.
\end{eqnarray}
The last term in the above Hamiltonian cannot be removed by near-identity 
transformations since it has non-zero mean with respect to $t$. However,
all other terms with rapidly oscillating coefficients can be killed.
Thus, choosing $W_{2}=\epsilon^{2}\int a_{-2}p_{2}\sin{\phi_{1}}$ 
we obtain the Hamiltonian
$H_{2}=\int \frac{p_{2}^{2}}{2}+\frac{\phi_{2x}^{2}}{2} +
\frac{1}{2}\epsilon^{2}\langle a_{-1}^{2} \rangle \sin^{2}{\phi_{2}} + 
\epsilon^{2}R + O(\epsilon^{2})$, where $\langle R \rangle = 0$. Finally,
taking $W_3=\epsilon^{3}\int  R_{-1}$ we obtain the Hamiltonian 
\begin{eqnarray}
H_3 = \int \left(\frac{p_3^{2}}{2}+\frac{\phi_{3x}^{2}}{2}+\frac{1}{2}
\epsilon^{2}\langle a_{-1}^{2} \rangle \sin^{2}{\phi_3}\right) 
+ O(\epsilon^{3})\;.
\label{finalh}
\end{eqnarray}
After rescaling $X=\epsilon x, \; T = \epsilon t,\;
P = 2\epsilon^{-1}p_3, \; \Phi = 2\phi_3$ in the equations of motion
corresponding to (\ref{finalh}), we obtain
\begin{eqnarray}
\left \{ \begin{array}{l}
\Phi_{T} = P + O(\epsilon^{2}) \\
P_{T} = \Phi_{XX} -  \langle a_{-1}^{2} \rangle \sin{\Phi} + O(\epsilon).
\end{array} \right.
\label{bigup}
\end{eqnarray}
The system (\ref{bigup}) is a slightly perturbed SGE
with $2\pi$-kinks as approximate solutions.
After rescaling back to variables $(\phi_3,p_3)$,
we obtain $\pi$-kinks as approximate solutions
$\phi_{3} \approx U(x,t)$, where
\begin{eqnarray}
U(x,t) = 2\arctan{\left[\exp{\left(
\epsilon \sqrt{\langle a_{-1}^{2}\rangle}
\frac{x-ct}{\sqrt{1-c^{2}}}\right)}\right]}\;.
\label{soliton}
\end{eqnarray}
Note, that by using the normal form technique, our equation (\ref{igor9}) has 
been brought to the form in which it explicitly represents 
a slightly perturbed SGE (\ref{bigup}).
Since we have employed only near-identical transformations,
the original equation (\ref{igor9}) should have solutions 
close to the solitary waves given by Eq. (\ref{soliton}).

\begin{figure}[h]
\vspace{1cm}
\hspace{-0.5cm}
\rightline{ \epsfxsize = 7.0cm \epsffile{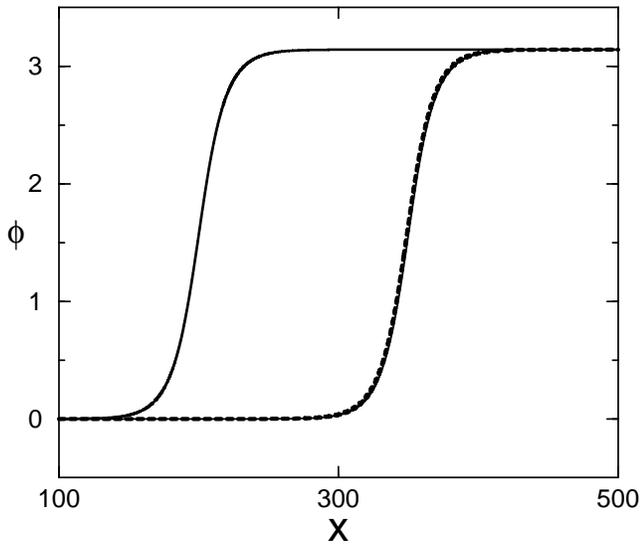}}
\caption{ The behavior of $\pi$-kink solution in the
PSGE (\protect \ref{igor9}). The thick solid line corresponds to the
initial profile given by (\protect \ref{fi_0}) with
$c = 1/2$, $a = \protect \sin(t/\epsilon)$.
The thin solid and dashed lines correspond to the
approximate solution (\protect \ref{soliton}) and the result of simulation,
respectively, at $t=300$. The analytical and numerical curves
almost coincide. The parameters of the simulations are
$\epsilon = 0.1$, time step $dt = 0.01$, mesh size $dx = 0.05$, system size
$L = 500$.
\label{fig1}}
\end{figure}

We have verified our results by the numerical simulations of PSGE
(\ref{igor9}), using the second-order leap-frog method.
To obtain initial conditions for the original variables,
$\phi(x,0), \, p(x,0)\,$, we start with the initial conditions
for the transformed variables, $\phi_3, \, p_3\,$,
generated from Eq. (\ref{soliton}) as $\phi_3(x,0)=
U(x,0)\,,\; p_3(x,0) = U_t(x,0)\,$.
Retracing our canonical transformations and keeping the lowest order terms only
we arrive at
\begin{eqnarray}
&&\phi(x,0) = U(x,0)\;, \nonumber 
\\
\label{fi_0}\\
&&p(x,0) = U_t(x,0) - \epsilon a_{-1}(0)\sin{U(x,0)}\;.
\nonumber
\end{eqnarray}
In Figure~\ref{fig1} we compare the results of the simulations (dashed
line) with the analytical solution (\ref{soliton}) (thin solid line).
One can see from the Figure a good quantitative agreement between the theory
and numerical simulations (the two curves in the Figure are almost
indistinguishable). We have also simulated PSGE (\ref{igor9}), starting
with initial conditions of shapes different from (\ref{fi_0}).
We have found that these solutions split into two linear
wave packets moving in
opposite directions with velocities $c=\pm 1$. This can be explained
by the closeness of PSGE with mean-zero excitation
to the linear wave equation, as was mentioned above.

\vspace{-1cm}

In terms of the original physical system described by Eq. (\ref{igor7}),
the $\pi$-kink solutions represent moving
true domain walls generated by a rapidly oscillating
external magnetic field.
Now we return from the dimensionless variables to the original variables
$\;t\,=\, \tilde{t}\,/\,C\,\sqrt{\nu H_0}\,,
\;x = \tilde{x}\,/\,\sqrt{\nu H_0}\;$,
to estimate the values of physical parameters,
which provide the formation of a realistic domain wall.
The lattice constant $a_0$ and energy constants
$J$ and $D$ are taken, according to Ref.~\cite{steiner},
as $a_0\,\sim\, 5\,$\AA, and $J\,\sim\, D\, \sim\, 10\,K$.
In dimensionless units the frequency $\tilde{\omega}$
of the field and the width of the domain wall $\Delta\tilde{x}$
are of order $\epsilon^{-1}\,$.
Then, for the field  amplitude $H_0\,\sim\,1\,G$ and
$\epsilon\,\sim\,0.1\div 0.01\,$, we obtain
$\omega\,\sim\,10^9\,Hz\,,\; \Delta x\,\sim\,10\,\mu m\,$.
Magnetic field $H_0\sin{\omega t}$, with such 
values of $H_0$ and $\omega$ can be created in an experiment,
to observe the predicted effect.

In summary, we have found $\pi$-kink solutions to
PSGE with a fast periodic mean-zero forcing. As applied to
quasi-one-dimensional ferromagnets with the only anisotropy of
an easy plane,
these solutions imply moving true domain walls, generated by
a rapidly oscillating magnetic field. Our theoretical results
are in a good quantitative agreement with numerical simulations
of the PSGE. These results are also applicable
to essentially two- and three-dimensional easy plane ferromagnets,
when plane front solutions are stable with respect to small
deformations. This problem, as well as the stability of $\pi$-kinks
in PSGE, will be addressed in future investigations.

We would like to thank S. Brazovsky, R. Camassa and V. Beilin
for fruitful discussions.


\vspace{-0.5cm}

\begin{references}

\vspace{-1.5cm}

\bibitem{scott} D. W. McLaughlin and A. C. Scott,
Physical Review {\bf A 18}, 1652 (1978).

\bibitem{mineev1} M. B. Mineev and V. V. Shmidt,
Sov. Phys. JETP {\bf 52}, 453 (1980).

\bibitem{frank} F. C. Frank and J. H. van der Merwe, Proc. R. Soc. London,
Ser. {\bf A 198}, 205 (1949).

\bibitem{bishop2} M. Rice, A. R. Bishop, J. A. Krumhansl, and S. E. Trullinger,
Phys. Rev. Lett. {\bf 36}, 432 (1976).

\bibitem{leung1} K. M. Leung et al., Phys. Rev. {\bf B 21}, 4017 (1980).

\bibitem{leung2} K. M. Leung, Phys. Rev. {\bf B 27}, 4017 (1983).       

\bibitem{mikeska1} H. J. Mikeska, J. Phys. {\bf C 11}, L29 (1978).

\bibitem{bishop1} A. R. Bishop and T. F. Lewis, J. Phys. {\bf C 12}, 3811 
(1979).

\bibitem{abdul} F. H. Abdullaev and P. K. Habibullaev, {\it Dynamics of 
solitons in non-homogeneous condensed media} (Tashkent, FAN, 1986) (in Russian) 

\bibitem{kirova} V. M. Yeleonosky and N. N. Kirova, Zh. Eksp. Teor. Fiz.
{\bf 75}, 2210 (1978).

\bibitem{steiner} J. K. Kijems, M. Steiner, Phys. Rev. Lett. {\bf 41}, 1137 
(1978); M. Steiner, Z. Phys. {\bf B 53}, 117 (1983).

\bibitem{jough} L. J. De Jough, Phys. Rev. Lett. {\bf 47} 1672 (1981).

\bibitem{kosevich} A. M. Kosevich, B. A. Ivanov, and A. S. Kovalev,
Phys. Rep. {\bf 194}, 117 (1990).

\bibitem{kivshar1} Yu. S. Kivshar, N. Gronbech-Jensen, and R. D. Parmentier,
Phys. Rev. {\bf E 49}, 4542 (1994).


\bibitem{kivshar2} Yu. S. Kivshar, N. Gronbech-Jensen, and M. R. Samuelsen,
Phys. Rev. {\bf B 45}, 7789 (1992).


\bibitem{bishop3} J. Wysin, A. R. Bishop, and P. Kumar, 
J. Phys. {\bf C 17}, 5975 (1984).

\bibitem{landau1} L.D. Landau, M. Lifshitz, {\em Mechanics}, (Pergamon Press, 
Oxford 1960).

\bibitem{arnold} V. I. Arnold, {\em Geometrical methods in the theory of 
ordinary differential equations}, (Springer-Verlag 1983).

\bibitem{neistadt} A. I. Neistadt, J. Appl. Math. Mech. 45(1981), no. 1,
58-63(1982).

\bibitem{levi} M. Levi, Curvature effects in averaging with applications, 
preprint (1997).

\end{references}
\end{document}